\documentclass[%
 reprint,
 amsmath,amssymb,
 aps,
]{revtex4-2}

\usepackage{graphicx}
\usepackage{xcolor}
\usepackage{tikz}
\usepackage{dcolumn}
\usepackage{bm}
\usepackage{graphicx}
\usepackage{nicefrac}

\usepackage[colorlinks=true,linkcolor=darkblue,urlcolor=darkblue,filecolor=darkblue,citecolor=darkblue,linktocpage=true]{hyperref}

\definecolor{lightgray}{gray}{0.9}
\definecolor{darkblue}{rgb}{0.2, 0, 0.8}

\renewcommand{\[}{\begin{equation}\begin{aligned}}
\renewcommand{\]}{\end{aligned}\end{equation}}

\def\mT{\mathcal{T}}

\def\hbar{\bar{h}}

\begin{document}

\preprint{APS/123-QED}

\title{New Near Extremal Black Holes and Love Symmetry}

\author{Alfredo Guevara}
    \altaffiliation{School of Natural Sciences, Institute for Advanced Study, Princeton NJ 08540, USA}
\author{Uri Kol}%
    \altaffiliation{Center of Mathematical Sciences and Applications, Harvard University, MA 02138, USA}


\date{\today}

\begin{abstract}
Rotating and charged black holes are known to exhibit remarkable properties close to extremality, including emergent hidden symmetries and holographic duality to 2D theories. In this note, we introduce a new class of near-extremal black holes living in $(2,2)$ signature, strongly resembling the Lorentzian ones but with an exact integrable structure $SL(2,\mathbb{R})\times SL(2,\mathbb{R})$. The exterior of the black hole is a self-dual solution with a photon ring. It develops an infinite near-horizon throat described by the Eguchi-Hanson instanton. An exact version of the so-called Love symmetry controls perturbations on the throat. Furthermore, the normalizable spectrum of this black hole provides quasinormal modes in Lorentzian signature.

\end{abstract}

\maketitle


\section{\label{sec:level1}  Introduction}

Most astrophysical black holes in the universe are expected to lie near extremality. A theoretical characterization of such cold black holes is that they develop an infinitely large throat near their horizon, where their symmetries are enhanced \cite {Bardeen:1999px,Guica:2008mu,Bredberg:2009pv,Castro:2010fd}. For the astrophysical case of a near-maximally rotating or even near-maximally charged black holes, the throat features a symmetric $AdS_2 \times S^2$ region and its perturbations decouple from its exterior. In recent years, this observation has led to surprising consequences, including their holographic description in terms of conformal field theories 
\cite{Iliesiu:2020qvm,
  Iliesiu:2022kny,
  Liu:2018hzo,
  Maldacena:2016upp,
  Mertens:2017mtv,
  Stanford:2017thb,
  Saad:2019lba,
  Iliesiu:2019xuh,
  Mertens:2020hbs}, in turn motivating observational considerations \cite{Casals:2016ues,Gelles:2025accumulation}. 

While in Lorentzian signature the no-hair theorem rules out other possible constructions of extremal black holes, it is interesting to consider alternatives with direct implications on Lorentzian measurements. A possibility arises in Kleinian (i.e. split) signature, where the addition of a NUT charge $N$ leads to the famous Taub-NUT solution. It shares many aspects with the charged RN solution, including bifurcate horizons, leading to its recent reinterpretation as a Kleinian black hole \cite{Crawley:2021auj}. A particularly interesting case occurs when the mass is equal to the NUT charge $M=N$, in which case the metric is a solution of Self-Dual (SD) gravity \cite{Guevara:2023wlr,guevara2024kol,inspirehep:???}. It is also know that another limiting case $M\to N$ leads to an interesting self-dual space with a conical singularity, covered by the so-called Eguchi-Hanson (EH) instanton. They have recently received a significant amount of attention \cite{inspirehep:2733846,Fenwick:2025fgg,inspirehep:1693554,Franchetti:2023wcw} due to newly surfacing connections to double copy, celestial holography, $w$-algebras and integrable structures.

Integrable structures are usually viewed as too simplistic to describe phenomenology. In this note, however, we will show that the suitable approached self-dual black hole shares many of the key features of its Lorentzian (near extremal) counterpart, while at the same time providing an integrable structure. Furthermore, we determine its quasi-normal spectrum and argue that it is controlled exactly by a novel near horizon symmetry, recently observed in other contexts, dubbed Love symmetry \cite{inspirehep:2147166,inspirehep:1849541,CombLoveNumbers}.

\section{Near the Kleinian horizon}
We start by considering the Kleinian Taub-NUT metric (we follow the conventions of \cite{Guevara:2023wlr})
\[
ds^{2}=(r^{2}-N^{2})\left(\frac{dr^{2}}{\Delta}+f\sigma_0^2 - \sigma_{+}\sigma_{-}\right),
\label{eq:ehlm}
\]
with lapse $\Delta=r^{2}-2Mr+N^{2}$ and \textit{warp function}
\[
f=\frac{4N^{2}\Delta}{(r^{2}-N^{2})^{2}}.
\]
The $AdS_3$ 1-forms are given by
\begin{align}\label{eq:1fom}
\sigma_{0} & =\frac{dt}{2N}-(\cosh\theta-\zeta)d\phi,\\
\sigma_{+}\sigma_{-} & =d\theta^{2}+\sinh^{2}\theta d\phi^{2}.
\end{align}
This is the $a\to0$ limit of the well-known Kerr Taub-NUT (KTN) metric. For the spinning case, $a\neq0$, the Hawking temperature
is $T_{H}=\frac{M\Omega}{2\pi a}\mathcal{T}$ where 
\[
\mathcal{T}=\frac{r_{+}-r_{-}}{2M}.
\]
Removing the horizon redshift factor $\frac{M\Omega}{2\pi a}$ \cite{Guevara:2023wlr},
we see that the dimensionless temperature $\mathcal{T}$ is well defined
as $a\to0$ and it measures the distance from extremality of the Taub-NUT metric
\[
\mT=\frac{\sqrt{M^{2}-N^{2}}}{M}.
\]
Analytically continuing to Lorentzian signature ($t\to it$ and $N\to iN$) we see that
\[
\mT=\frac{\sqrt{M^{2}+N^{2}}}{M}\geq1 .
\]
Equivalently, eliminating $N$ in favor of $\mathcal{T}$ in \eqref{eq:ehlm}, the region $\mT \geq 1$ describes a Lorentzian slice \footnote{Related to the usual Lorentzian slice by $\theta \to i\theta$.} and $\mathcal{T}<1$ describes a Kleinian slice. Only the later requires an identification $t \sim t +\beta (\mathcal{T})$ \cite{Guevara:2023wlr}. As we are ultimately interested in Lorentzian physics this will not be imposed at this point.

In this paper we consider cold Black Holes with $\mathcal{T}\ll1$ which
have emergent symmetries. For that purpose we introduce the following radial coordinate
\[\label{xrad}
x=\frac{r-M}{M},
\]
in terms of which we have
\[
\Delta & =M^{2}(x^{2}-\mathcal{T}^{2}), \\
f & =\frac{4(1-\mathcal{T}^{2})(x^{2}-\mathcal{T}^{2})}{(2x+x^{2}+\mathcal{T}^{2})^{2}}\approx\frac{4(x^{2}-\mathcal{T}^{2})}{(2x+x^{2}+\mathcal{T}^{2})^{2}},
\]
(and setting $N\to M$ in $\sigma_0$, eq. \eqref{eq:1fom}). This is our definition of \textit{Near SD Black Hole}. It develops an infinite
throat in the region $x\sim\mathcal{T}$ analogous to the usual near-horizon geometry. The geometry is obtained by scaling 
\[
    x\to\lambda x, \qquad
    \mT\to\lambda\mT,
\]
and taking the leading order in $\lambda$. In this scaling limit we see that the warp function goes to
\[
f\to f_{EH}=1-\frac{\mT^{2}}{x^{2}},
\]
which is the the warp factor of the Eguchi-Hanson (EH) instanton. In this limit the metric \eqref{eq:ehlm} takes the form
\[\label{eq:ehlm2}
ds^2_{EH}=
2M^{2}x\left(\frac{dx^{2}}{x^{2}-\mathcal{T}^{2}}+f_{EH}\sigma_{0}^{2}-\sigma_{+}\sigma_{-}\right).
\]
The usual EH metric  
\cite{Berman:2018hwd} is recovered
by setting $x=\rho^{2}/8M^{2}$ and $\mathcal{T}=a_0^{2}/8M^{2}$ in \eqref{eq:ehlm2}, where $a_0$ is the EH parameter.

To analyze the region away from the throat we simply take $\mathcal{T}\to\lambda\mathcal{T}$and
keep $x$ fixed. This gives
\[
f\to f_{SDTN}=\frac{4}{(x+2)^{2}}
\]
which is the known warp function of SD Taub-NUT (SDTN). In that case we have
\[
ds^{2}_{SDTN} = M^{2}x(x+2)\left(\frac{dx^{2}}{x^{2}}+f_{SDTN}\sigma_{0}^{2}-\sigma_{+}\sigma_{-}\right),
\label{eq:sdtn2}
\]
which is the usual SDTN if we set $x=\rho/M$. The extra factor of $\frac{1}{\lambda}$
in (\ref{eq:ehlm2}) as compared to (\ref{eq:sdtn2}) indicates that
the mass of a probe particle will get redshifted near horizon.

\begin{figure}[htbp]
\centering
\begin{tikzpicture}[thick,line cap=round,line join=round,xscale=0.5]
\fill[green!18] (-6,1.6) -- (6,1.6) -- (7,4) -- (-5,4) -- cycle;
\shade[top color=green!45, bottom color=blue!65, samples=200, smooth]
  (-3,3)
    .. controls (-2.3,3)  and (-2.0,2.6) .. (-1.2,1.8)
    .. controls (-0.6,1.1) and (-0.2,1.0) .. (0,1.0)
    .. controls (0.2,1.0)  and (0.6,1.1) .. (1.2,1.8)
    .. controls (2.0,2.6)  and (2.3,3)   .. (3,3)
    arc[start angle=0,end angle=180,x radius=3,y radius=0.55]
  -- cycle;
\draw (0,3) ellipse [x radius=3, y radius=0.55];
\draw[<->] (-3,3.6) -- (3,3.6);
\node at (0,3.85) {$\sim \mathcal{T}^{2/3}$};
\node at (0,1.8) {EH};
\node[below] at (0,1.0) {$x=0$};
\draw (-6,1.6) -- (6,1.6) -- (7,4) -- (-5,4) -- cycle;
\node at (4.8,3.35) {SDTN};
\end{tikzpicture}
\caption{Topology of the near extremal BH in $r,t$ (each point contains a $(\theta,\phi)$ hyperboloid). The Kleinian solution is regular at the horizon $x=0$. A throat develops at $\mathcal{T}\ll1$ and shrinks to a point for $\mathcal{T}=0$ (this is the hydrogen atom studied in \cite{Guevara:2023wlr}). In the Lorentzian case the throat has the topology of $AdS_2$ as expected. }
\label{fig:sdtn-funnel)}
\end{figure}
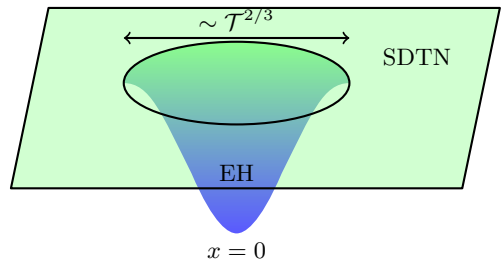

Note that if we take $\mathcal{T}\to0$ in (\ref{eq:ehlm2}) or $x\to0$
in (\ref{eq:sdtn2}) we get the ``strict horizon'' metric, that
is nothing but (thermal) flat space
\[
ds^{2}\to2M^{2}x\left(\frac{dx^{2}}{x^{2}}+\sigma_{0}^{2}   -\sigma_{+}\sigma_{-}\right),
\label{eq:hza}
\]
and which is where the Love symmetry applies according to \cite{Guevara:2023wlr}.

\section{The Photon ring and resonances}

In the Schwarzchild geometry, long lived resonances are developed inside
the photon ring but can tunnel through the centrifugal barrier. What is the fate of such resonances in
near SD black holes? As we approach the limit we expect they are controlled by critical behavior and its emergent symmetries. A first hint is the size of the photon ring which
scales, as we will show, with a critical exponent $2/3$,
\[\label{eq:scaq}
r_{c}\sim\mT^{\nicefrac{2}{3}}\,,\quad \mathcal{T}\ll 1\,
\]
(this is the universality class of the fractional quantum Hall effect \cite{Subramanyan_2021})
We further show that the putative CFT associated with the critical behavior
has a symmetry group $SL(2,\mathbb{R})\times SL(2,\mathbb{R})$. One of these 
$SL(2,\mathbb{R})$ is associated to the angular problem. The other is associated to a wave equation on $AdS_2$ whose solutions, from the 4d Lorentzian perspective, are precisely quasi-normal modes.

\begin{figure}
    \centering
    \includegraphics[width=\linewidth]{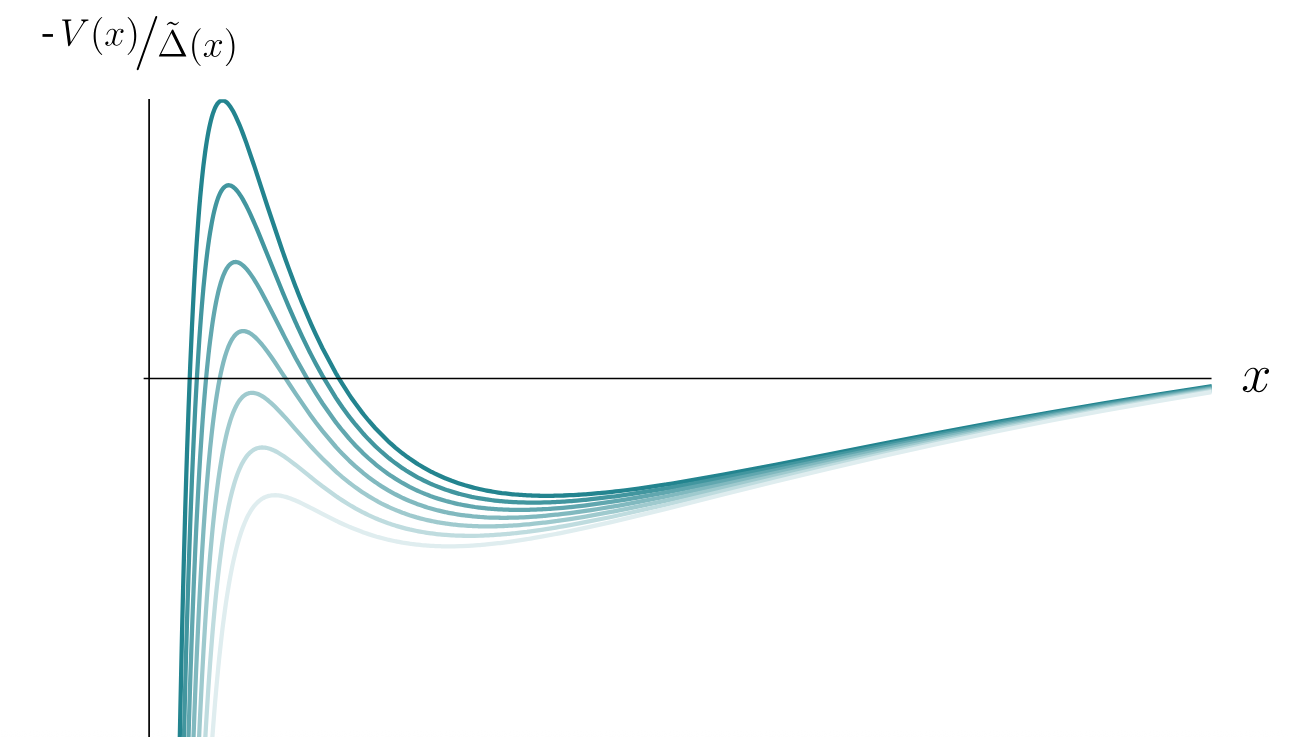}
    \caption{Plot of $\frac{-V(x)}{\Delta(x)}$ for low (dark) to high (light) temperatures. We have continued $\omega \to i\omega$ to reach the Lorentzian slice, resembling the Scharwzschild potential. The unstable maximum corresponds to the location of the photon ring, which approaches the horizon as $\mT \rightarrow 0$ (and becomes unbounded at the same time).}
    \label{fig:potential}
\end{figure}

From \eqref{eq:scaq} we see that in the strict $\mT=0$ limit
the photon ring shrinks and resonances decouple. Equivalently, the strength of the centrifugal barrier becomes infinite and one recovers
the Coulomb problem (see figure \ref{fig:potential}). The spectrum takes discrete degenerate eigenvalues,
just as for the hydrogen atom \cite{Guevara:2023wlr}. On the other hand, for $\mT\ll1$
but non-zero, resonances have an exponentially small chance of escaping
the photon ring.
In the next sections we will compute that decay width and argue that it matches the spectrum of quasi-normal modes (QNM) of the black hole.

To simplify the analysis, we shift the radial coordinate in \eqref{xrad} as
\[
x\to x+\mathcal{T},
\]
such that now $x=\frac{r-r_{+}}{M}$ and $\Delta\to \tilde{\Delta}=x(x+2\mT)$. The shifted radial coordinate has the
advantage that for Schwarzchild $x=0$ is the horizon and $x=1$ is
the photon ring. The wave equation in Klein signature is now given by
\[
\partial_{x}(\tilde\Delta(x)\partial_{x}\psi)+V(x)\psi & =0,
\]
where the potential is
\[
V(x)  &= -J^{2}+(2M\omega)^{2}-\frac{H(x)^{2}}{\tilde{\Delta}}M^{2}\omega^{2}, \\
H(x)  &=-1+\mathcal{T}^{2}+(1+x+\mathcal{T})^{2},
\]
and where $J^{2}$ is the Casimir of the group of isometries of \eqref{eq:ehlm}.
The ratio $V(x)/\tilde\Delta(x)$ is in fact the radial potential of a null geodesic, and we plot it in figure \ref{fig:potential} for illustration. The critical point associated
to the photon ring is then given by the loci
\[\label{loci}
V(x_{c})=V'(x_{c})=0,
\]
which is solved by \footnote{This formula is also derived from the full potential $V(x)$ without neglecting $\mathcal{T}^2$ terms. Thus also holds near $\mathcal{T}\to 1$.}
\[\label{criticalLocation}
x_{c} & =\mathcal{T}\left(-1+\left(\frac{\mathcal{\mathcal{T}}_{+}}{\mathcal{\mathcal{T}}_{-}}\right)^{\nicefrac{1}{3}}+\left(\frac{\mathcal{\mathcal{T}}_{-}}{\mathcal{\mathcal{T}}_{+}}\right)^{\nicefrac{1}{3}}\right),
\]
where
\[
\mT_{\pm} & =\sqrt{1\pm \frac{N}{M}},
\]
(such that $\mT_+\mT_-=\mT$ and
$\frac{\mT_+}{\mT_-}=\frac{\mT}{1+\sqrt{1-\mT^{2}}}$). Note that \eqref{loci} also implies a condition on the frequency, but the value of the critical frequency will not be important here.

From \eqref{criticalLocation} we conclude that 
\[
x_{c} & \sim\mathcal{T}&\quad \text{when}\quad&\qquad\mathcal{T\sim}1, \\
x_{c} & \sim\mathcal{T}^{\nicefrac{2}{3}}&\quad \text{when}\quad&\qquad\mathcal{T}\ll1.
\]
The first line is the behavior associated with the Schwarzchild BH,
for which $x_{c}=1$ or $r_{c}=3M$. The second line is obtained by
allowing $\mT<1$ which goes into the Kleinian section. Since
$x_{c}$ determines the peak of the centrifugal barrier, one can see
that as $\mathcal{T}\to0$ the barrier sticks to the origin but becomes
unbounded at the same time, hence leading to the Coulomb potential.

\section{Quasi-Normal Modes}

The SD black hole is integrable \cite{guevara2024kol} and consequently its spectrum of fluctuations is described by normal modes with real frequencies. Away from the SD point this is not true anymore, and the resulting spectrum of quasi-normal modes develops an imaginary part. In this section, we compute the spectrum of QNM for the near-SD class of black holes introduced earlier.

\begin{figure}
    \centering
    \includegraphics[width=1\linewidth]{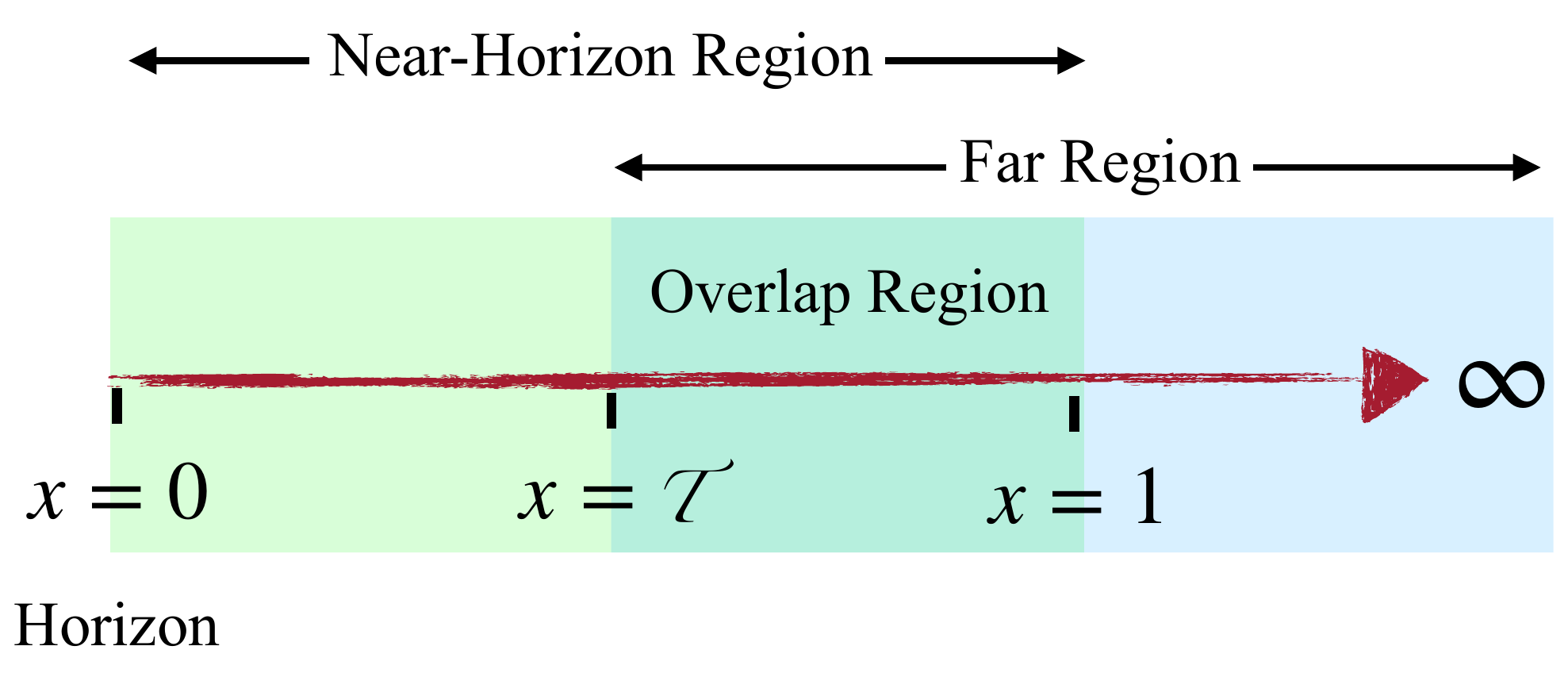}
    \caption{The matching procedure between solutions of the near-horizon and far regions, for near-SD black holes with $\mT\ll 1 $. Note that in this regime the photon ring is located deep in the near-horizon region. However, as $\mT\rightarrow 1$ two things happen: the overlap region shrinks to zero size and the photon ring moves towards the outer boundary of the near-horizon region.}
    \label{fig:matching}
\end{figure}

Since the SD black hole is integrable, the wave equation on this background is solved exactly in terms of simple polynomials \cite{Guevara:2023wlr}. Beyond the SD point, this is not true anymore and therefore we will construct the solution using a matched asymptotic expansion valid for near-SD black holes $\mT \ll 1$. We match two solutions which are valid in different regions of spacetime, which is possible due to the scaling \eqref{eq:scaq} of the matching region. The near-horizon region is where $x\ll 1$ and the far region is where $x \gg \mT$. Boundary conditions are imposed at the horizon using the near-horizon solution. For $\mT$ small enough the matching region is $\mT \ll x \ll 1$. See figure \ref{fig:matching} for illustration of the procedure.

\textit{Near Zone Equations}. First note that in the near-horizon region $x \ll 1$ the metric takes the form of the Eguchi-Hanson instanton \eqref{eq:ehlm2}. The wave equation in this near-horizon near-SD throat region is exactly given by the Casimir operator of a hidden $SL(2,\mathbb{R})$
which is indeed the so-called Love symmetry of \cite{Guevara:2023wlr}
\[
\left(L_{0}^{2}-\frac{L_{+}L_{-}+L_{-}L_{+}}{2}\right)\psi=J^{2}\psi,
\label{eq:sfds1}
\]
here the new $SL(2,\mathbb{R})$ generators are given by 
\[
L_{0} & =i \beta\partial_{t},\\
L_{+} & =ie^{-it/\beta}(\sqrt{\tilde\Delta}\partial_{x}-i\beta\partial_{x}\sqrt{\tilde{\Delta}}\partial_{t}),\\
L_{-} & =ie^{it/\beta}(\sqrt{\tilde\Delta}\partial_{x}+i\beta\partial_{x}\sqrt{\tilde\Delta}\partial_{t}),
\]
with $\beta=2M$.  The generators can be identified with the isometries of an (Euclidean) $AdS_2$ space,
\begin{equation}
    ds_2^2=\frac{\tilde{\Delta}}{2\mathcal{T}^2}dt^2+\frac{2M^2}{ \tilde{\Delta}}dx^2
\end{equation}and in fact we use this as a guiding principle to determine the spectrum:
\[
L_{0} & =2M\omega  \, , \\ 
L^{2} &= J^2 = \hbar (\hbar-1),
\]
We now show that \eqref{eq:sfds1} has a single solution (rather than two), which corresponds to a quasi-normal mode. We have
\[
\psi_{\text{NH}}=P_{\hbar-1}^{-2M\omega}(1+\frac{x}{\mathcal{T}})e^{i\omega t}\times Y_{\hbar-1,m}^{2M\omega}(\theta,\phi)
\]
which is diagonal in both $SL(2,\mathbb{R})$. Here $Y^q_{\hbar-1,m}(\theta,\phi)=F^q_{\hbar-1,m}(\theta)e^{im\phi}$ are the standard monopole harmonics and we will denote by
\[
u_{\text{NH}}(x):=P_{\hbar-1}^{-2M\omega}(1+\frac{x}{\mT}) \sim x^{M\omega}
\]
the radial component of the wavefunction. This is the solution corresponding to the incoming boundary conditions at the horizon in the Lorentzian sense. From the Kleinian perspective, however, this is the regular solution as $x\to 0$, with the other `shadow' solution blowing up as $x^{-2M\omega}$.

In \cite{Guevara:2023wlr}, it was found that \eqref{eq:sfds1} is the exact wave equation at the horizon (\ref{eq:hza}).

Here we have extended this result to the infinite near-horizon throat of near-SD black holes. In this sense, the putative CFT is described by $SL(2,\mathbb{R})\times SL(2,\mathbb{R})$
primaries. Crucially, until this point we do not specify the representation $\bar{h}$. In the Kleinian space we use discrete series \cite{Atanasov_2021} and analytically continue the angular momentum  to Lorentzian signature later. 

\textit{Far Zone and Matching}. In the far region, the metric reduces to that of SDTN \eqref{eq:sdtn2}. Solutions of the wave equation on this background are given by \cite{Guevara:2023wlr}
\begin{equation}
    \psi_{\textrm{far}}=u_{\text{far}}(x) \,\,e^{i\omega t}\times Y_{\hbar-1,m}^{2M\omega}(\theta,\phi)
\end{equation}
Here the radial function is the confluent hypergeometric function,
\[
&u_{\text{far}}(x) = \\
&+c_1 \times e^{M\omega x}x^{\hbar-1} {}_1F_1  \left( \hbar-2M\omega;2\hbar; -2M\omega x \right) \\
&+ c_2 \times  \left( \hbar \rightarrow  1-\hbar \right).
\]
For the exact SDTN background, this far region extends all the way to the horizon, where the regular boundary conditions select only one of the modes. Indeed, this space only has `normal' modes and no absorption. However, this is not the case for near-SD black holes, where the far region approximation is not valid close to the horizon. In that case, the ratio between the modes will be determined by the matching to the near-horizon solution in the overlap region.

 The matching region is described by $\mT \ll x \ll 1$. To proceed, we first expand the near-horizon solution around $x \gg \mT$, which results in
\[\label{eq:matchingNH}
u_{\text{NH}}(x) & \sim x^{\hbar-1}+\mathcal{T}^{2\hbar-1}\alpha_{\hbar}\frac{1}{x^{\hbar}},
\]
where
\[
\alpha_{\hbar} & =\frac{\Gamma(-\hbar+\frac{1}{2})\Gamma(\hbar+2M\omega)}{2^{2\hbar-1}\Gamma(\hbar-\frac{1}{2})\Gamma(-\hbar+1+2M\omega)}.
\]
Then, we expand the far region solution around $x \ll 1$
\[\label{eq:matchingFar}
u_{\text{far}}(x) \sim c_1 \, x^{\hbar-1}
+c_2 \, \frac{\Gamma(2\hbar-1)}{\Gamma(\hbar-2M\omega)} \frac{1}{x^{\hbar}}.
\]
Matching \eqref{eq:matchingNH} and \eqref{eq:matchingFar} we can now fix the ratio
\[\label{FarSolCof}
&\frac{c_2}{c_1}=\mathcal{T}^{2\hbar-1}\alpha_{j}\frac{\Gamma(\hbar-2M\omega)}{\Gamma(2\hbar-1)}
\\
&=\mathcal{T}^{2\hbar-1}
\frac{\Gamma(-\hbar+\frac{1}{2})\Gamma(\hbar+2M\omega)\Gamma(\hbar-2M\omega)}{2^{2\hbar-1}\Gamma(\hbar-\frac{1}{2})\Gamma(2\hbar-1)\Gamma(-\hbar+1+2M\omega)},
\]
thus completely determining the solution corresponding to the incoming boundary conditions, to leading order in the expansion parameter $\mT$.

\textit{Quasinormal spectrum}.  The spectrum of QNM is determined by imposing boundary conditions at infinity, in addition to the boundary conditions at the horizon, turning the variational problem non-Hermitian. Consequently, the energy spectrum of fluctuations is complex.

The asymptotic form of the solution at $x\rightarrow \infty$ is given by the following superposition
\[\label{asySol}
u_{\text{far}} &\sim 
B_{\text{out}}e^{- M \omega x_{\star}} 
+
B_{\text{in}}e^{+ M \omega x_{\star}},
\]
where
\[
x_{\star} = x+ 2 \log x
\]
is the tortoise coordinate and the amplitudes take the form
\[
B_{\text{out}}&=
c_1 \, (-2M\omega)^{-\hbar-2M\omega}
\frac{\Gamma(2\hbar)}{\Gamma(\hbar-2M\omega)},\\
B_{\text{in}} &=
c_1 \,  (2M\omega)^{-\hbar+2M\omega} \frac{\Gamma(2\hbar)}{\Gamma(\hbar+2M\omega)}
\\
&+c_2 \,  (-2M\omega)^{\hbar-1+2M\omega}.
\]
Note that in Lorentzian signature the first and second terms in \eqref{asySol} correspond, respectively, to out-going and in-coming waves.

Now the boundary conditions that we wish to impose at infinity are that in Lorentzian signature there is only an out-going mode, namely we wish to set to zero the amplitude of the in-coming mode
\[\label{QNMeq}
B_{\text{in}} = 0.
\]
In Kleinian signature that translates into the requirement that the divergent term at infinity is set to zero.
This condition will fix the spectrum of frequencies.

Plugging \eqref{FarSolCof} into \eqref{QNMeq} we find the following equation for $\omega$
\[\label{QNMeq2}
\frac{\left(i M \omega\right)^{1-2\hbar}e^{-2iM\omega \pi}  }{i \mT ^{2\hbar-1} \alpha_{\hbar}}=
 \frac{\Gamma(\hbar+2M\omega)\Gamma(\hbar-2M\omega)}{\Gamma(2\hbar-1)\Gamma(2\hbar)}.
\]
To solve this equation, recall that we expand in small $\mT$ around the integrable configuration. We can therefore parametrize the spectrum as
\[\label{omegaAnsatz}
\omega _{n\hbar}=  \frac{n}{2M}+ \frac{1}{2M} \mT^{2\hbar-1} \eta_{n\hbar},
\]
where the first term is the intergrable spectrum \cite{Guevara:2023wlr}, and $\eta_{n\hbar}$ is dimensionless and of order $\sim 1$. The power law behavior of $\mT$ will be justified shortly. Now note that within all the Gamma functions in equation \eqref{QNMeq2} there is only one that we should worry about -
\[
\Gamma(\hbar-2M\omega) = \Gamma(  -m- \mT^{2\hbar-1} \eta_{n\hbar}   ),
\]
where
\[
m\equiv n-\hbar = 0,1,2,\dots
\]
is a non-negative integer. Note that when $\mT \rightarrow 0$ this Gamma function diverges as
\[
 \Gamma(  -m- \mT^{2\hbar-1} \eta_{n\hbar}   ) = \frac{(-1)^m}{m!  \, \mT^{2\hbar-1} \, \eta_{n\hbar}} + \dots
\]
where the dots stand for finite or vanishing terms. Notice that the divergence of this Gamma function will exactly cancel the factor $\mT^{2\hbar-1}$ in \eqref{QNMeq2}, therefore justifying the $\mT$ dependence that we chose in \eqref{omegaAnsatz}. At the end we find that
\begin{equation}\label{eq:mainnm}
\eta_{n\hbar} =\left( \frac{n}{2} \right)^{2\hbar-1}
\frac{\Gamma(-\hbar+\frac{1}{2})\left(\frac{\Gamma(n+\hbar)}{\Gamma(n-\hbar+1)}\right)^2}{\Gamma(\hbar-\frac{1}{2})\Gamma(2\hbar-1)\Gamma(2\hbar)}.
\end{equation}
This is our final result for the QNMs of near SD black holes. Closely resembling expressions have appeared in the context of small AdS Black Holes \cite{Cardoso_2004,Festuccia:2008zx,Berti_2009,Morgan:2009vg}, which is analogous to the extremality condition here. Next we proceed to interpret this result in the Lorentzian picture.

\section{Lorentzian Tunneling}

As a final discussion we attempt to interpret the result \eqref{eq:mainnm} as long-lived QNM in Lorentzian signature. The idea is as follows: We can access Lorentzian signaure via analytic continuation $\omega \to i\omega $ and $n\to i n$, and argue that the second term in \eqref{omegaAnsatz} corresponds to the imaginary part of the Lorentzian frequency \footnote{More precisely, in the discrete series $\hbar=k+1/2 \in \mathbb{Z}+1/2$ and $\omega\to\frac{n}{2M}+i\frac{(-1)^k }{2M} \mathcal{T}^{2k}\eta$}. 
The fact that it scales as a power law in $\mathcal{T}$ suggests that these are long-lived modes for an extremal black hole.

To confirm this we compute the decay rate for a particle to tunnel from the Eguchi-Hanson throat into the asymptotic Self-dual region.
We will show that it agrees with our results for the QNM in the regime $\mathcal{T}\ll 1$. The estimation is done in the eikonal regime for
$J,M\omega\gg1$. Let $p_x$ be the Kleinian radial momentum,  then the transition probability is given by
\[
\Gamma=\exp\left(i\oint p_{x}\,dx\right)=\exp\left(2\int_{x_{-}}^{x_{+}}\sqrt{-\frac{V(x)}{\tilde\Delta(x)}}\,dx\right),
\]
The second equality is evaluated in the Lorentzian slice $\omega^2<0$, for which the potential is depicted in figure \ref{fig:potential}, and then rotated back to match \eqref{eq:mainnm}. The Lorentzian interpretation is a space-like
particle tunneling from the near horizon region at $x_{-}$ to
the outer region at $x_{+}$. For $\mathcal{T\ll}1$ we expect
\begin{equation}
    \Gamma \propto \mathcal{T}^{2\hbar-1}\approx\mathcal{T}^{2J}\,.
\end{equation}
with $J^2=h(h-1)$ i.e. $J\approx \hbar-\frac{1}{2}$. Let us compute this explicitly. In the limit we have
\begin{align}
x_{+} & \approx\frac{\sqrt{J^{2}-4M^{2}\omega^{2}}}{iM\omega}-2-\mathcal{T}, \\
x_{-} & \approx\mathcal{T}\left(\frac{\sqrt{J^{2}-4M^{2}\omega^{2}}}{J}-1\right).
\end{align}
Thus confirming the endpoints occur in the far and near-horizon regions, respectively. 
To approximate the integral we will consider the splitting $x\in[x_{+},x_{c}]$
and $x\in[x_{c},x_{+}]$. Here $x_c$ is defined by $V'(x_{c})=0$, resulting in the following expression \footnote{but $V(x_{c})\neq0$ as opposed to Section 1, since we allow motion
for the virtual particle.} 
\[
x_{c}\approx\mathcal{T}\left(\frac{\sqrt{J^{2}-8M^{2}\omega^{2}}}{J}-1\right).
\]
As the first interval has contracting length $x_{-}-x_{c}=\mathcal{O}(\mathcal{T})$
it will be enough to consider
\begin{align}
&2\int_{x_{c}}^{x_{+}}\sqrt{-\frac{V(x)}{\tilde\Delta(x)}}\,dx  =2\int_{x_{c}+\mathcal{T}}^{x_{+}+\mathcal{T}}\sqrt{-\frac{V(x-\mathcal{T})}{\tilde\Delta(x-\mathcal{T})}}\,dx \\
 & \approx2 M\omega\int_{\epsilon}^{\frac{\sqrt{J^{2}-4M^{2}\omega^{2}}}{i M\omega}-2}\sqrt{x^{2}+4x+\frac{J^{2}}{M^{2}\omega^{2}}}\,\,\frac{dx}{x}.
\end{align}
In the first equality we shifted $x\to x-\mathcal{T}$, returning
to the variable used in the first section. Recall that then the
integrand is quadratic in $\mathcal{T}$, thus we can evaluate it
at $\mathcal{T}=0$. This is the approximation of the second line.
In the lower limit of this integral we find a logarithmic regulator
with 
\[
\epsilon=\mathcal{T}\sqrt{1-\frac{8M^{2}\omega^{2}}{J^{2}}},
\]
accounting for the actual throat contribution. The integral is elementary
(e.g. can be evaluated by residues) and becomes
\begin{align}
2J(1-2\log{J}+\log{M\omega\epsilon)}+(2M\omega+J)\log(M\omega+J/2) \nonumber\\
-(2M\omega -J)\log(M\omega-J/2)
\end{align}
which can be approximated using the Stirling
formula as $M\omega,J\gg1$, leading to
\begin{equation}\label{eq:gamp}
\Gamma \approx 
\frac{\left(M\omega\epsilon\right)^{2J}}{J!^4}\frac{\Gamma(1+M\omega+J/2)}{\Gamma(1+M\omega-J/2)}\frac{\Gamma(1-M\omega+J/2)}{\Gamma(1-M\omega-J/2)}
\end{equation}
Reinserting $M\omega\approx n/2, J\approx \hbar-1/2 $ we obtain $\Gamma\propto \left(\frac{n\mathcal{T}}{2} \right)^{2\hbar-1}$ as expected from \eqref{eq:mainnm}. It would be nice to infer the numerical factors relating \eqref{eq:mainnm} to \eqref{eq:gamp} along the lines of the formulae in \cite{Berti_2009,Festuccia:2008zx}.

\section{Discussion}

We have introduced a new class of near extremal Black Holes. They exhibit analogous properties to the more familiar Reissner-Nordstrom solutions, but they instead live naturally in Kleinian signature. Most of the analysis is simplified by the emergent $SL(2,\mathbb{R})\times SL(2,\mathbb{R})$ symmetry at critical temperature, and subleading corrections in $\mathcal{T}$ are expected to spontaneously break it. This provides strong clues of a holographic dual as in previous examples \cite{Maldacena:2016upp,Guevara:2023cqecc, Kapec:2023logKerr, Detournay:2025photonrings}.

In \cite{guevara2024kol} we studied the complementary problem of scattering between Self-Dual solutions in the probe limit, which is controlled by quadratic generators extending the Laplace-Runge-Lenz vector. The symmetry emerging here, near the horizon, appears simpler and linearly realized. It would be interesting to extend it to the black hole two-body problem. 

In general self-dual black holes behave as `non-linear gravitons' and have integrable interactions \cite{Penrose:1976nonlinear_graviton}. It would be interesting to constrain such using the methods of \cite{Lupsasca_2025_NoLoveBH}, which has shown integrability reduces certain modes of the wave equation to a flat-space equation.

\begin{acknowledgments}
We thank Juan Maldacena, Lionel Mason, Atul Sharma and David Skinner for useful discussions. We specially thank Huy Tran for collaboration during earlier stages of this project.

UK is supported by the Center for Mathematical Sciences and Applications at Harvard University. AG acknowledges the Roger Dashen membership at the IAS, and additional support from DOE
grant DE-SC0009988.
\end{acknowledgments}

\bibliographystyle{apsrev4-2}
\bibliography{apssamp}

\end{document}